\documentclass[aps,pra,twocolumn,groupedaddress]{revtex4}
\usepackage{graphicx}
\newcommand{\beq}{\begin{eqnarray}}
\newcommand{\eeq}{\end{eqnarray}}
\begin{document}

\title{Probing the evolution of Stark wave packets by a weak half cycle pulse}

\author{H. Wen, S. N. Pisharody, J. M. Murray, and P. H. Bucksbaum}
\affiliation{FOCUS Center, Department of Physics, University of
Michigan, Ann Arbor, MI 48109-1120}
\date{\today}
\begin{abstract}
We probe the dynamic evolution of a Stark wave packet in cesium using weak half-cycle pulses
(HCP's). The state-selective field ionization(SSFI) spectra taken as a function of HCP delay reveal
wave packet dynamics such as Kepler beats, Stark revivals and fractional revivals. A
quantum-mechanical simulation explains the results as multi-mode interference induced by the HCP.

\end{abstract}
\maketitle

\section{Introduction}

Recent publications have shown that Rydberg states in an atom can be used as data storage registers
for quantum information processing~\cite{AhnScience2000}.  A significant limitation is that
relatively few states couple directly to the atomic ground state via allowed transitions, so that
the full range of Rydberg state quantum numbers cannot participate in quantum information
operations.  Quantum control methods have been proposed to extend the data registers to include
angular momentum states~\cite{WenPRA03}. To utilize this approach for information storage, the
angular momentum content of high-$\ell$ Rydberg wave packets must be measured with reasonable
fidelity. Here we show how to approach this using electromagnetic half-cycle pulses (HCP's) and
state-selective field ionization (SSFI).

Several experiments have utilized correspondence between classical and quantum systems to relate
classical dynamics to the control and manipulation of a quantum system~\cite{BromagePRL99,
ArboPRA03, StokelyPRA96}. Meanwhile, other experiments have utilized a direct quantum mechanical
approach to understand and control the quantum properties of the wave packets~\cite{WeinachtPRL98,
AhnPRL01, AhnPRA02}. In either the classical or the quantum mechanical context, the efficient
detection of a wave packet is always a challenge. Several detection methods have been used with
varying degrees of success, for example, short pulse pump-probe ionization\cite{NoordamPRA89},
bound state-state interferometry \cite{NaudeauPRA97}, and time-resolved streak camera detection of
ionization\cite{LankhuijzenPRL96}. Detection of Stark wave packet dynamics has also been
successfully demonstrated using HCP's~\cite{CampbellPRA98, CampbellPRA99, RamanPRA97}. In those
previous experiments, the detection relies on the fact that an HCP induces a change of electron's
total energy, $\Delta E=\vec{p}\cdot \vec{Q}+\vec{Q}^{2}/2$, to ionize the electrons. Here
$\vec{Q}$ represents the momentum transferred to electrons from the HCP and $\vec{p}$ is the
initial momentum of the electron. In these experiments, the ionization of the atoms by HCP-assisted
transfer of momentum to the electrons is used to retrieve the classical dynamics of the electronic
wave packet. However, these methods of detection are insensitive to the effect of the HCP on the
bound-state population distribution in the wave packet. In the context of quantum information
processing, control over the amplitudes and phases of each of the states in the wave packet is
essential for the manipulation of quantum information.

In this work, we use a weak HCP to probe the evolution of a Stark wave packet. The population
redistribution in the wave packet due to the HCP~\cite{TielkingPRA95} is detected by SSFI. Kepler
beats, angular momentum revivals and fractional revivals are all clearly observed in the HCP-delay
dependent SSFI spectra. In contrast to time-delay spectroscopy~\cite{CRamanPRL1996} and impulsive
momentum retrieval~\cite{JonesPRL96}, the strength of the HCP's used in this experiment are much
smaller than that required to directly ionize the atoms.

\section{Experimental Procedure}

A thermal beam of cesium atoms in a uniform electric field, $F\hat{z}$, are driven from the $6s$
ground state to the $7s$ launch state via two-photon excitation by the focused 1079nm output of a
Ti:Sapphire-laser-pumped OPA.  A shaped, amplified, ultrafast laser pulse excites Stark wave
packets with $m_\ell=0$ in the range of $24 \leq n^* \leq 28$.

A weak HCP with a peak field of about 1 kV/cm and a pulse width of 400 fs is applied along the
direction of the static electric field following the creation of the Stark wave packet. The HCP is
created by illuminating a high-voltage-biased GaAs photoconductive switch with the 800nm, 50 fs
output of the Ti:Sapphire laser. The HCP applies an impulse of 0.002 a.u.(atomic units:
$e=m_e=\hbar=1$) to the electrons. This redistributes the bound state populations and produces
multi-mode interference depending on the relative phases of the states. About $ 10 \mu s$ after
this impulse, SSFI is used to detect the energy level composition of the wave packet.

In the absence of an HCP, SSFI is insensitive to the evolution of the angular momentum composition
of the states in the wave packet. A weak HCP differentiates the SSFI spectra based on the
time-dependent angular momentum composition of the wave packet. We can then use the SSFI spectra of
the HCP-kicked Stark wave packet to calibrate its angular momentum composition.

\begin{figure}
\centering\rotatebox{0}{\includegraphics[width=3.3in]{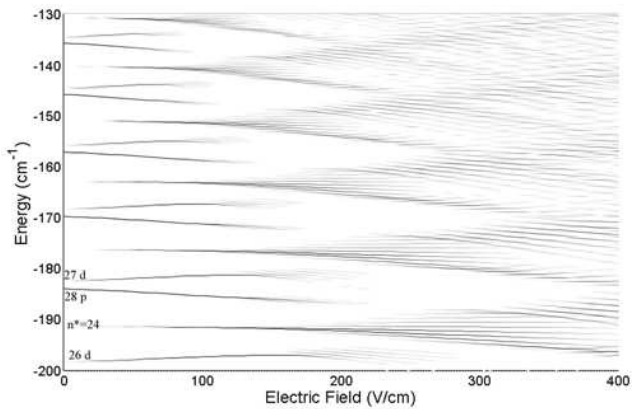}} \caption{ \label{fig1}The Stark
energy levels for Cs($m=0$) in the range of $24 \leq n^{*} \leq 28$. The darkness of the lines
indicate the photo-excitation probability at the applied external electric fields. }
\end{figure}

\section{Data Analysis}

\begin{figure}
\centering \rotatebox{0}{\includegraphics[width=3.3in]{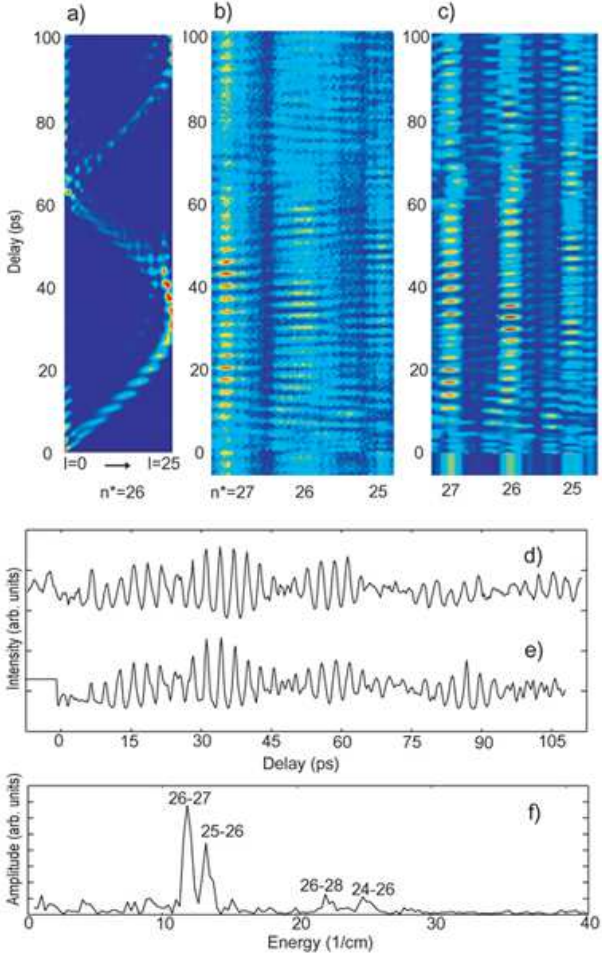}} \caption{\label{fig2}a) The
angular momentum components of the $n^*\sim26$ state in the Stark wave packet as a function of time
following wave packet excitation; b) The state-resolved, HCP-induced, time-dependent spectrum,
known as a quantum carpet. The ionization signal(color) of four manifolds($25\leq n^*\leq 28$) is
shown as a function of HCP delay in a static electric field $F=160V/cm$; c) The calculated quantum
carpet for b); d) The ionization signal for $n^*\sim 26$ as a function of HCP delay; e) The
calculated ionization signal for d); f) The Fourier transform of d). }
\end{figure}

A wave packet excited from the 7s launch state in Cs to a set of excited Rydberg states in the
presence of an external electric field has p-state character immediately after its excitation.
Angular momentum, however, is not a good quantum number for Rydberg states in an electric field.
The coupling of angular momentum in the different energy levels of the Stark wave packet results in
the time-dependent variation of the angular momentum of the wave packet \cite{WenPRA03}. This is
shown in Fig.\ref{fig2}(a). In the presence of the static electric field, the angular momentum of
the $n$ manifold in the Stark wave packet precesses from low $\ell$ to high $\ell$ and back over
the course of a Stark period of $\sim \frac{2 \pi}{3Fn}$. The angular momentum evolution is
understood classically as a precession of the electron orbit under the influence of the external
electric field.

Fig.\ref{fig2}(b) shows the SSFI spectrum plotted as a function of the delay of the HCP kick
relative to the creation of the Stark wave packet for the case of an external electric field of 160
V/cm. The HCP redistributes the population in the different manifolds depending on the
instantaneous angular momentum composition and phase of the states in the wave packet. The spectrum
at the bottom of the graph (for delay $\leq$ 0) corresponds to the initial population distribution
in the Stark wave packet before the HCP interaction(Fig.\ref{fig1}). For the first 10-15ps
following the excitation of the wave packet, the HCP redistributes population predominantly into
low angular momentum states. The SSFI spectrum changes significantly in this regime due to the
large quantum defects for the $p$ and $d$ states compared to the other states in the manifold. At
later delays of the HCP, there are periodic modulations in intensity of the SSFI spectrum for each
of the states in the wave packet. A line-out of the intensity variation in the spectrum
corresponding to the $n^*\sim 26$ state is shown in Fig.\ref{fig2}(d). We find that the spectrum
returns to its low angular momentum form at HCP delays around 64 ps. This corresponds to the Stark
period ($\frac{2 \pi}{3Fn}$) of a wave packet centered around $n^*\sim 26$ at a field of 160 V/cm.
The envelope of intensity modulations over one Stark period also displays minima in intensity at
times 26 ps and 48 ps. This temporal structure corresponds to the fractional revival periods,
$\tau_{f}=2\pi n^{4}/3$ \cite{YeazellPRL90}. Around these visibility minima, the Kepler beats
double in frequency~\cite{YeazellPRA91} due to a splitting of the wave packet into two parts.

The Fourier transform of Fig.\ref{fig2}(d) is shown in Fig.\ref{fig2}(f). The peaks correspond to
the energy differences between the state $n^*\sim~26$ and its neighboring states. The weak HCP acts
to couple neighboring $n$ states and to produce interference between different $n$ manifolds in the
Stark wave packet. The two strongest frequency components are attributed to the energy differences
of its neighboring states $\Delta E_{26,27}=11.8cm^{-1}$, and $\Delta E_{25,26}=13.3cm^{-1}$. The
two weaker peaks correspond to the energy differences of its next-neighbor states $\Delta
E_{26,28}=22.4cm^{-1}$, and $\Delta E_{24,26}=24.8cm^{-1}$. The deviation of $\Delta E_{24,26}$
from the zero-field value of $28.2cm^{-1}$ is due to manifold mixing at $F=160V/cm$ so that the HCP
couples the Stark states at the edge of the $n^*\sim 26$ manifold instead of in the middle.

\begin{figure}
 \centering \rotatebox{0}{\includegraphics[width=3.3in]{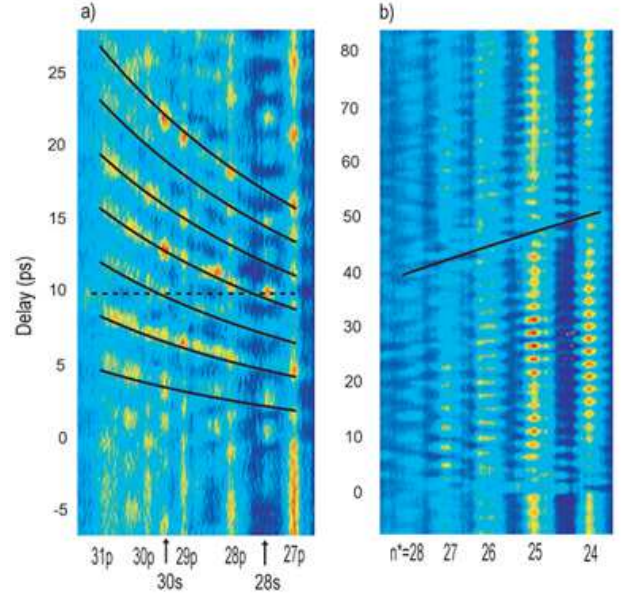}} \caption{\label{fig3}a)The quantum carpet for
$F=80V/cm$. The dark lines represent the successive energy-dependent Kepler periods
($\tau_{Kepler}=2\pi n^3$); b) The quantum carpet for $F=240V/cm$. The dark line represents the
Stark revival time $\tau_{Stark}=2\pi /3Fn$.}
\end{figure}

At low electric fields, the $p$ and $d$ states are completely separated in energy from the higher
angular momentum components of the Stark manifolds. The pattern for $F=80 V/cm$ in
Fig.\ref{fig3}(a) shows dominant population shifts from low energy states to high energy states as
a function of the HCP delay. This ``quantum carpet'' pattern fits well with the Kepler periods for
corresponding $n$-states in the wave packet\cite{AhnPRL01}. The dark lines in the figure are
plotted to represent the expected times for Kepler revivals ($\tau_{Kepler}=2\pi n^3$) for each
state in the wave packet. These are found to correspond very closely to the observed maxima in the
HCP-assisted population redistribution in the wave packet. In Fig.\ref{fig3}(b), where $F=240V/cm$,
a calculated line indicates the expected times for angular momentum revivals ($\tau_{Stark}=2\pi
/3Fn$). This matches the repetition of the extended features in the SSFI spectrum seen in the
experiment.

Fig.\ref{fig3}(a) also demonstrates the use of a weak HCP to selectively populate particular low
angular momentum states by an intelligent choice of HCP delay. The HCP arriving at the interaction
region 11 ps after the creation of the Stark wave packet in an external field of 80 V/cm
selectively populates the $28 s$ and the $30 s$ states while removing population from the other
states.

\section{Numerical Simulation}

We have performed numerical simulations to gain further insight
into our experimental results. We model the HCP as an impulsive
momentum kick to the electronic Stark wavepacket\cite{AhnPRA02}.

The interaction Hamiltonian of the HCP is written as
 \beq
 H(t)=-\vec{\mathcal{F}}(t)\cdot\vec{r}.
 \eeq where we have used atomic units throughout. The
result of the HCP interaction with the wave packet can then be
written as
 \beq
  |\Psi_{f}(t)\rangle=e^{i\vec{Q}\cdot \vec{r}}|\Psi_{i}(t)\rangle
 \eeq
where $|\Psi_{i,f}(t)\rangle$ are the wave packets in the basis of the Stark Hamiltonian before and
after the interaction with the HCP. The integral, $\vec{Q}=-\int \vec{\mathcal{F}}(t')dt'$
represents the total momentum transferred to the electron. The $\hat{z}$-polarized HCP is modeled
as a unitary operator $e^{iQz}$ in the impulse approximation. Immediately before the arrival of the
HCP, $\Psi$ is written as \beq |\Psi_{i}(t)\rangle=\sum_{nk} c_{nk}e^{-i\omega_{nk}t}|nk\rangle
\eeq

After the interaction with the HCP, the $\Psi$ becomes \beq
|\Psi_{f}(t)\rangle=e^{iQz}|\Psi_{i}(t)\rangle=\sum_{nk,n'k'}
\mathcal{M}_{n'k',nk}c_{nk}e^{-i\omega_{nk}t}|n'k'\rangle, \eeq where \beq
\mathcal{M}_{n'k',nk}=\langle n'k'
 |e^{iQz}|nk\rangle
\eeq
 represents the HCP effect. The coupling matrix elements are calculated in the field-free
$n\ell m$-basis and subsequently transformed into the $nkm$(Stark)-basis using the transformation
matrix that diagonalizes the Stark Hamiltonian, $H=\frac{p^2}{2}-1/r+Fz$. The strongest HCP
coupling occurs between the states in adjacent manifolds rather than between states within the same
manifold. For the HCP strengths used, the dominant coupling occurs for states separated by one or
two manifolds. This explains the observed energy level contributions to the dynamics in
Fig.\ref{fig2}(e). Multi-mode interference due to HCP coupling among neighboring manifolds enhances
the Kepler-period modulations in the observed SSFI spectra.

The calculations are performed over a range of energies corresponding to $10\leq n^* \leq 40$. The
state distribution in the final wave packet after the HCP interaction is mapped from its energy,
$E$, to the expected ionization field, ($F_{i}=E^2/4$), for comparison to the experimental SSFI
spectrum. The results of our simulations with an external electric field of 160 V/cm is shown in
Fig.\ref{fig2}(c) and the line-out corresponding to the $n^*\sim 26$ state is shown in
Fig.\ref{fig2}(e). The results of our simulations are in excellent agreement with the experiment.

\section{Discussion and Conclusions}

An HCP acting on an electronic wave packet changes the electron energy depending on its
instantaneous linear momentum along the direction of the HCP. This feature has enabled experiments
in the past that have used directed HCP's to map the time-resolved momentum distribution of the
electron in its orbit\cite{CampbellPRA99,RamanPRA97}. When the Stark wave packet evolves to high
angular momentum, the corresponding classical orbits of the electron are along the lines of
constant longitude on the surface of a sphere where the poles are aligned in the direction of the
electric field. In the semi-classical picture, the high angular momentum $m=0$ electron wave packet
is well localized radially and oscillates in latitude between the poles~\cite{RamanPRA97}. Such a
wave packet has a linear momentum aligned along the electric field at times separated by a Kepler
period when it is localized at angles corresponding to the equator on the sphere. The wave packet
also has the least linear momentum in the direction of the HCP at times when it is localized close
to the poles of the sphere. This maximal variation in linear momentum character and therefore
maximal variation in energy transfer to the electrons can be seen in our data and simulations to
correspond to the greatest contrast in Fig.\ref{fig2}(d) and Fig.\ref{fig2}(e) at times around 33
ps.

In summary, we employ a weak HCP as a coherent redistribution operator to probe the Stark wave
packet. The quantum carpet that we obtain reveals three characteristic times: the Kepler beats, the
Stark revivals, and the fractional revivals, and agrees well with their expected $n$-dependence.
Our simulations using an impulse model for the HCP are in excellent agreement with the experimental
results. The use of a weak HCP to differentiate the instantaneous angular momentum composition of a
Stark wave packet could, in the future, allow us to use SSFI as a single-shot detector for angular
momentum states.

\section{Acknowledgements}
It is a pleasure to thank C. Rangan and F. Robicheaux for useful discussions. This work was
supported by the National Science Foundation under Grant No.9987916 and the Army Research Office
Grant No.DAAD 19-00-1-0370.

\bibliography{StarkWpInt}

\end{document}